\documentclass[aps,preprint]{revtex4}
\usepackage{natbib}
\usepackage{graphicx,subfigure,float,graphics}
\usepackage{epsfig}

\begin{document}

\title{Isovector and isoscalar proton-neutron pairing  in $N>Z$ nuclei}

\author{D. Negrea$^a$, P. Buganu$^a$, D. Gambacurta$^b$ and 
N. Sandulescu$^a$\footnote{corresponding author, email: sandulescu@theory.nipne.ro}}
\affiliation{
$^a$ National Institute of Physics and Nuclear Engineering, 
077125 M\v{a}gurele, Romania \\
$^b$ Extreme Light Infrastructure - Nuclear Physics (ELI-NP), 
National Institute of Physics and Nuclear Engineering, 077125 M\v{a}gurele, Romania}

\begin{abstract}
We propose a particle number conserving formalism for the treatment of
isovector-isoscalar pairing in nuclei with $N>Z$.   The ground state of the
pairing Hamiltonian is described by a quartet condensate to which is appended
a pair condensate formed by the neutrons in excess. The quartets are built by
two isovector pairs coupled to the  total isospin $T=0$ and two collective
isoscalar proton-neutron pairs. To probe this
ansatz for the ground state we performed calculations for $N>Z$ nuclei with the valence
nucleons moving above the cores $^{16}$O, $^{40}$Ca and $^{100}$Sn. The calculations
are done with two pairing interactions, one state-independent and the other of zero 
range, which are supposed to scatter pairs in time-revered orbits. 
It is proven  that the ground state correlation energies calculated within this
approach are very close to the exact results provided by the diagonalization of the
pairing Hamiltonian. Based on this formalism we have shown that moving away of N=Z line, 
both the isoscalar and the isovector proton-neutron pairing correlations remain significant
and that they cannot be
treated accurately by models based on a proton-neutron pair condensate.

\end{abstract}

\maketitle

\section{\label{sec:intro}Introduction}

In spite of many years of theoretical and experimental studies, the role of neutron-proton 
pairing in nuclei is still a matter of debate (for recent reviews, see \cite{frauendorf,sagawa}).
One of the most debated issue is whether in nuclei the isoscalar (T=0) spin-triplet neutron-proton
pairs could form a "deuteronlike" pair condensate and if this condensate would coexist with the
condensates of spin-singlet isovector (T=1) pairs 
\cite{satula,terasaki,goodman99,goodman2001,gezerlis}.
Most of the studies have been carried out for heavy nuclei close to the N=Z line in which the
spin-triplet pairing is  expected to be stronger and less suppressed by the spin-orbit field. Some 
calculations predict that a spin-triplet phase might exist, alone or mixed with the isovector pairing 
phase, in the ground state of some nuclei \cite{goodman99,gezerlis}, but it is not clear yet  how much 
these predictions are affected by the employed approximations. On the experimental side, so far 
there is no clear evidence for the fingerprints of the proton-neutron pairing phase on 
measurable quantities \cite{frauendorf}.

The majority of the theoretical studies mentioned above have been done by treating the pairing
in the framework of the generalized Bogoliubov approach, as outlined many years ago by 
Goodman \cite{goodman_review}. This approach is very convenient because can treat all types of pairing
correlations, isovector and isoscalar, on an equal footing. It has however the drawback that does
not conserve exactly the particle number and the isospin. As noticed many years ago \cite{lane}, 
an isospin conserving theory can be formulated in terms of alpha-like 4-body structures (called
hereafter "alpha-like quartets" or, simply, "quartets") built by coupling two protons and two 
neutrons to total isospin T=0. One of the first attempts on this line was the treating of
the isovector pairing by a BCS-like state based on alpha-like quartets \cite{flower}. 
Similar BCS trial states have been employed to study the competition between pairing and
quartetting in nuclei \cite{bremond,zelevinsky}. Alpha-like quartets have been also used
to treat pairing in particle number conserving formalisms \cite{yamamura,dobes}. One of the first such formalisms,
employed for the treatment of isovector pairing, was based on quartets built by two neutrons and 
two protons sitting in the same single-particle orbit \cite{yamamura}. Since this formalism is using non-collective
quartets, its application to large systems is cumbersome. An alternative alpha-like quartet formalism, 
based on collective quartets, was proposed in 
Refs. \cite{qcm_nez,qcm_ngz}. In this formalism the ground state of isovector paring Hamiltonians is 
described  as a condensate of collective quartets in the case of N=Z nuclei \cite{qcm_nez}, 
and as a condensate of quartets to which it is appended a condensate of neutron pairs in 
the case of $N>Z$ nuclei \cite{qcm_ngz}. Recently this quartet condensation formalism (QCM) was
extended to treat both isovector and isoscalar pairing interactions in N=Z nuclei \cite{qcm_t0t1,qcm_odd}. 
The scope of this paper is to extend further this approach to $N >Z$ nuclei and to probe the
validity of the new approach for  isovector and isoscalar pairing Hamiltonians which
can be solved exactly by diagonalization.

\section{\label{sec:model}Formalism and calculation scheme}
\renewcommand{\theequation}{2.\arabic{equation}}

Since the present formalism is an extension of the  model introduced in Ref. \cite{qcm_t0t1},
for the sake of completeness we start by presenting shortly this approach. As in Ref. \cite{qcm_t0t1},
we consider systems formed by neutrons and protons moving in axially
deformed mean fields and interacting by isovector and isoscalar pairing forces which 
scatter pairs of nucleons in time-reversed single-particle states. These systems are
described by the Hamiltonian
\begin{eqnarray}
H&=&\sum_{i,\tau=\pm1/2}\varepsilon_{i\tau}N_{i\tau}+\sum_{i,j}V^{(T=1)}_{i,j}\sum_{t=-1,0,1}P_{i,t}^{\dag}P_{j,t}\nonumber\\
&&+\sum_{i,j}V_{i,j}^{(T=0)}D_{i,0}^{\dag}D_{j,0},
\label{ham}
\end{eqnarray}
where $\varepsilon_{i,\tau}$ are single-particle energies associated with the mean field of neutrons $(\tau=1/2)$ and protons $(\tau=-1/2)$ while $N_{i,\tau}$ are the particle number operators. The second term is
the isovector pairing interaction expressed by the isovector pair operators
$P_{i,1}^{\dag}=\nu_i^{\dag}\nu_{\bar{i}}^{\dag}$, $P_{i,-1}^{\dag}=\pi_i^{\dag}\pi_{\bar{i}}^{\dag}$, $P_{i,0}^{\dag}=(\nu_i^{\dag}\pi_{\bar{i}}^{\dag}+\pi_i^{\dag}\nu_{\bar{i}}^{\dag})/\sqrt{2}$. The third term is the
isoscalar pairing interaction and $D_{i,0}^{\dag}=(\nu_i^{\dag}\pi_{\bar{i}}^{\dag}-\pi_i^{\dag}\nu_{\bar{i}}^{\dag})/\sqrt{2}$ is the isoscalar pair operator. By $\nu^{\dag}_i$ and $\pi^{\dag}_i$ are denoted the
creation operators for neutrons and protons while $\bar{i}$ represents the time conjugate of 
the state $i$.
 
It is worth emphasising that the Hamiltonian (1), which is employed in many nuclear
structure calculations (e.g., see \cite{pacearescu} and the references quoted therein), 
treats the correlations associated to the pairs built on time-reversed axially deformed states.
As such, these pairs have $J_z=0$ but not a well-defined angular momentum $J$. 

In most of the studies the Hamiltonian (1) is treated in BCS-like approximations based
on the generalized Bogoliubov transformation. An alternative approach, which conserves exactly the particle number and the
isospin, was proposed in Ref. \cite{qcm_t0t1} for the case of
even-even N=Z systems. In this approach, called quartet condensation model (QCM), the ground state of the Hamiltonian (1) is approximated by the trial state
\begin{equation}
|QCM\rangle=(A^{\dag}+\Delta_0^{\dag 2})^{n_q}|0\rangle,
\label{qcm}
\end{equation}
where $n_q=(N+Z)/2$ while $|0\rangle$ is the "vacuum" state represented by the nucleons which are 
supposed to be not affected by the pairing interactions (e.g., an even-even closed core). 
The operator $A^{\dag}$ is the isovector quartet built 
by two isovector non-collective pairs coupled to the total isospin $T=0$, i.e.,
\begin{equation}
A^{\dag}=\sum_{i,j}x_{ij}[P^{\dag}_i P^{\dag}_j]^{T=0}.
\end{equation}
Assuming that the mixing coefficients are separable, i.e., $x_{ij}=x_i x_j$, 
the isovector quartet takes the form
\begin{equation}
A^{\dag}=2\Gamma_{1}^{\dag}\Gamma_{-1}^{\dag}-(\Gamma_{0}^{\dag})^{2},
\label{isovq}
\end{equation}
where $\Gamma_{t}^{\dag}=\sum_{i}x_{i}P_{i,t}^{\dag}$ are collective pair operators for neutron-neutron
pairs ($t=1$), proton-proton pairs ($t=-1$) and proton-neutron pairs ($t=0$). The isoscalar degrees of freedom are described by the
collective isoscalar pair
\begin{equation}
\Delta_{0}^{\dag}=\sum_{i}y_{i}D_{i,0}^{\dag}.
\label{isosq}
\end{equation}

The trial state (2.2) is called a quartet condensate. The term condensate has here the same meaning as in
the case of pair condensate: a state obtained by acting many times with the same operator on a
"vacuum" state.

In what follows we extend this approach to even-even systems with $N > Z$ (the case $N<Z$ is
treated in the same manner). As in the QCM approach presented above, by N and Z we denote the
numbers of neutrons and protons moving above a self-conjugate core which plays the role
of the reference (vacuum) state. To describe the ground state of the systems with $N > Z$
we use the following ansatz : (a) we assume that the protons together with an equal number of neutrons
are forming a 4-body condensate with the same structure as in Eq.(2.2); (b) we assume that the
neutrons in excess are forming a pair condensate which is appended to the 4-body condensate.
The trial state which corresponds to these assumptions is given by
\begin{equation}
|QCM\rangle=(\tilde{\Gamma}_1^{\dag})^{n_N} (A^{\dag}+\Delta_0^{\dag 2})^{n_q}|0\rangle,
\label{totalwf}
\end{equation}
where $n_N=(N-Z)/2$ gives the number of neutron pairs in excess while $n_{q}=(N+Z-2n_N)/4$ 
denotes the maximum number of quartets which can be formed with Z protons. The extra neutrons
are represented by the collective neutron pair
\begin{equation}
\tilde{\Gamma}_{1}^{\dag}=\sum_{i}z_{i}P_{i,1}^{\dag}.
\end{equation}
As can be seen, the structure of the extra pairs, expressed by the mixing amplitudes,
is different from the structure of the neutron pairs which enter in the definition
of the isovector quartet (2.4).

It is worth mentioning that  in the particular case when the isoscalar pairs are absent, 
the state (2.6) is the ansatz employed in Ref. \cite{qcm_ngz} for the description of 
the ground state of $N>Z$ systems interacting by an isovector pairing interaction.
 
The state (2.6) has a very complicated structure when it is expressed in terms of pairs.
Thus, replacing the quartet operator by (2.4) one can see that the state (2.6) is a superposition
of pair condensates, each of them formed by various types of pairs. Among these terms of special
interest are the following ones:
\begin{equation}
|C_{iv}\rangle=(\tilde{\Gamma}_{1}^{\dag})^{n_N}(\Gamma_0^{\dag 2})^{n_q}|0\rangle,
\label{aproxwf1}
\end{equation}
\begin{equation}
|C_{is}\rangle=(\tilde{\Gamma}_{1}^{\dag})^{n_N}(\Delta_0^{\dag 2})^{n_q}|0\rangle.
\label{aproxwf2}
\end{equation}
As can be seen, in the first (second) state it is supposed that the proton-neutron
correlations are described by a condensate of isovector (isoscalar) proton-neutron pairs.
The validity of these assumptions will be tested below against the full ansatz (2.6).

The trial state (2.6) depends on the mixing amplitudes of the collective pair operators.
They are determined variationally by minimizing the average of the Hamiltonian under the
normalisation condition imposed to the trial state. The average of the Hamiltonian and
the norm are evaluated using the method of reccurence 
relations based on the auxiliary states
\begin{equation}
|n_{1}n_{2}n_{3}n_{4}n_{5}>=\Gamma_{1}^{\dag n_{1}}\Gamma_{-1}^{\dag n_{2}}\Gamma_{0}^{\dag n_{3}}\Delta_{0}^{\dag n_{4}}\tilde{\Gamma}_{1}^{\dag n_{5}}|0>.
\end{equation}
Compared to the case of even-even N=Z systems described in Ref. \cite{qcm_t0t1}, 
these auxiliary states contain in addition the pair corresponding to the extra neutrons. 
This fact makes the recurrence relations much more complicated than for N=Z systems.
Moreover, the variational calculations are also more difficult because the number of 
variational parameters to be determined increased by $25\%$.
Hence, although the extension of the QCM state from (2.2) to (2.6) appears simple
from formal point of view, the calculations with the extended trial state is a difficult task.
In order to optimize the numerical calculations, for the particular systems treated in the
next section we have derived analytically the average of the Hamiltonian and the norm  
by employing symbolic computing algorithms. In this way the calculations can be performed
much faster, comparable to the BCS-like calculations.

\section{Results}
\renewcommand{\theequation}{3.\arabic{equation}}

To probe the accuracy of the approach presented above we use
the same examples as in Refs. \cite{qcm_ngz,qcm_t0t1}. Namely, we consider three
sets of nuclei with the valence neutrons and protons moving above the 
cores $^{16}$O, $^{40}$Ca and $^{100}$Sn. We start by the even-even N=Z systems 
obtained  by adding to each core one, two and three quartets, which are described
by the trial state (2.2). Then, on the top of these N=Z nuclei we add up to three 
neutron pairs; these $N>Z$ systems we describe by the trial state (2.6).
The nucleons are supposed to move in the lowest 10 single-particle states above the 
closed cores mentioned above. These states are generated by axially-deformed Skyrme-HF
calculations performed for the $N=Z$ nuclei. In the mean field calculations we have 
employed the Skyrme functional SLy4 \cite{sly4} and we have neglected the Coulomb interaction.
What remains to be chosen are the pairing interactions. How to fix these interactions
for nuclei with neutrons and protons in the same valence shell is not clear 
established, especially for the case of isoscalar pairing force. 
The simplest interaction which is usually taken in the isovector pairing
channel is a state-independent force. Here we have chosen such an interaction of strength
$V_1=-24/A$, where $A$ is the atomic mass of the nucleus. For the isoscalar interaction
we use the same force as in the isovector channel but of different strength, i.e.,
$V_0= w V_1$, where $w$ is a scaling factor. For the latter many values
have been employed in the literature, ranging  from $w=1.5$ \cite{sagawa,gezerlis} to values
smaller than one \cite{bentley}. To cover these situations, here we have chosen two values, 
$w=1.2$ and $w=0.8$. The first value we have employed for $sd$-shell nuclei and the latter for 
the heavier nuclei. We have made this choice because it is expected that in $pf$-shell nuclei
the isoscalar pairing interaction is more suppressed than in $sd$-shell nuclei due to the
spin-orbit splitting. 

In addition to the state-independent interactions mentioned above, we consider also
a zero-range delta interaction $V^T(r_1,r_2)=V_0^T\delta(r_1-r_2) \hat{P}^T_{S,S_z}$, 
where $\hat{P}^T_{S,S_z}$ is the projection operator on the spin of the pairs, i.e., 
$S=0$ for the isovector (T=1) force and  $S=1,S_z=0$ for the isoscalar (T=0) force. 
For the strengths $V_0^{T}$ we have chosen the values employed in Ref. \cite{qcm_odd}, 
which provide a reasonable description of the lowest $T=1$ and $T=0$ states in odd-odd N=Z nuclei. These values are $V_0^{T=1}=465$ MeV fm$^{-3}$
and $V_0^{T=0}= w V_0^{T=1}$, where $w=1.6$ for $sd$-shell nuclei and $w=1.0$ for the
heavier nuclei. 

\begin{figure}[h]
\begin{center}
\includegraphics[width=1.00\textwidth]{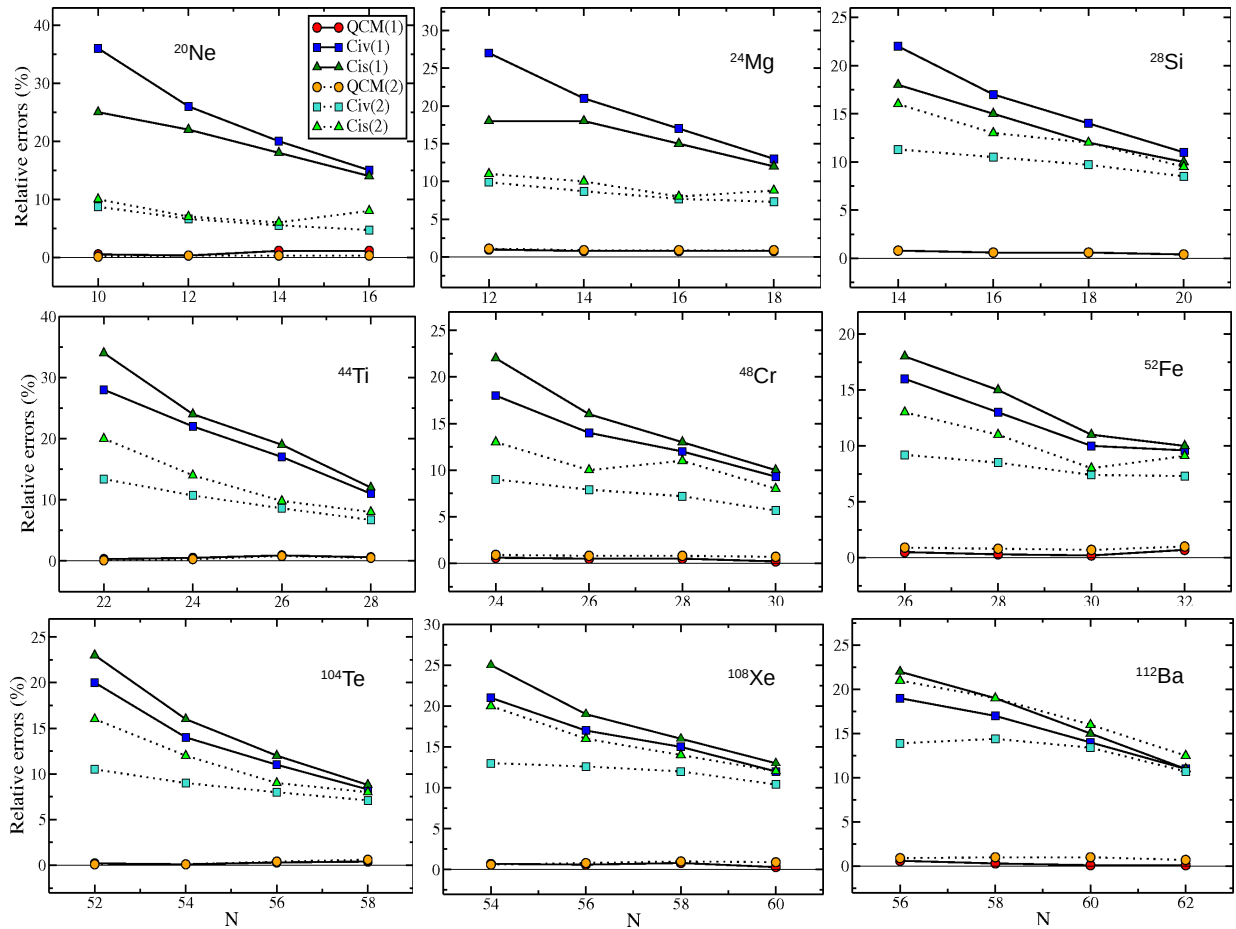}
\end{center}
\caption{The errors for the energy correlations as a function of neutron number for various isotopes. 
The results correspond to the QCM state (Eq. 2.6) and to the approximations $C_{iv}$ (Eq. 2.8) 
and $C_{is}$ (Eq. 2.9). The labels 1 and 2 in the brackets refer to the results obtained with the
state independent force and zero range force, respectively. The errors are calculated relative to the 
exact results.}
\label{fig1}
\end{figure}

With the two pairing forces  we have tested the ansatz (2.6) for the ground state energy
of Hamiltonian (1) considering the $N>Z$ systems mentioned above. To evaluate the accuracy
of the approach we have analysed the ground state correlation energies defined by $E_{corr}=E_0-E$,
where $E$ is the ground state energy and $E_0$ is the energy in the absence of the 
interactions. The correlation energies are compared to the exact values obtained by diagonalizing
the Hamiltonian (1). The errors, respecting to the exact results, are shown in Fig. 1. One can observe
that for all the systems the errors are small, under $1\%$, which demonstrates that the QCM ansatz (2.6) is describing very well the ground state pairing correlations.

In Fig. 1 are shown also the errors corresponding to the pair condensates given by Eqs. (2.8, 2.9). 
It can be seen that the errors corresponding to these trial states are much larger. They are the
largest for the N=Z nuclei and then they decrease for the systems with  extra neutrons. 
These results indicate that going off the N=Z line there is not a fast transition
towards a pure condensate of proton-neutron pairs, of isovector or isoscalar kind.

\begin{figure}[h]
\begin{center}

\includegraphics[width=1.00\textwidth]{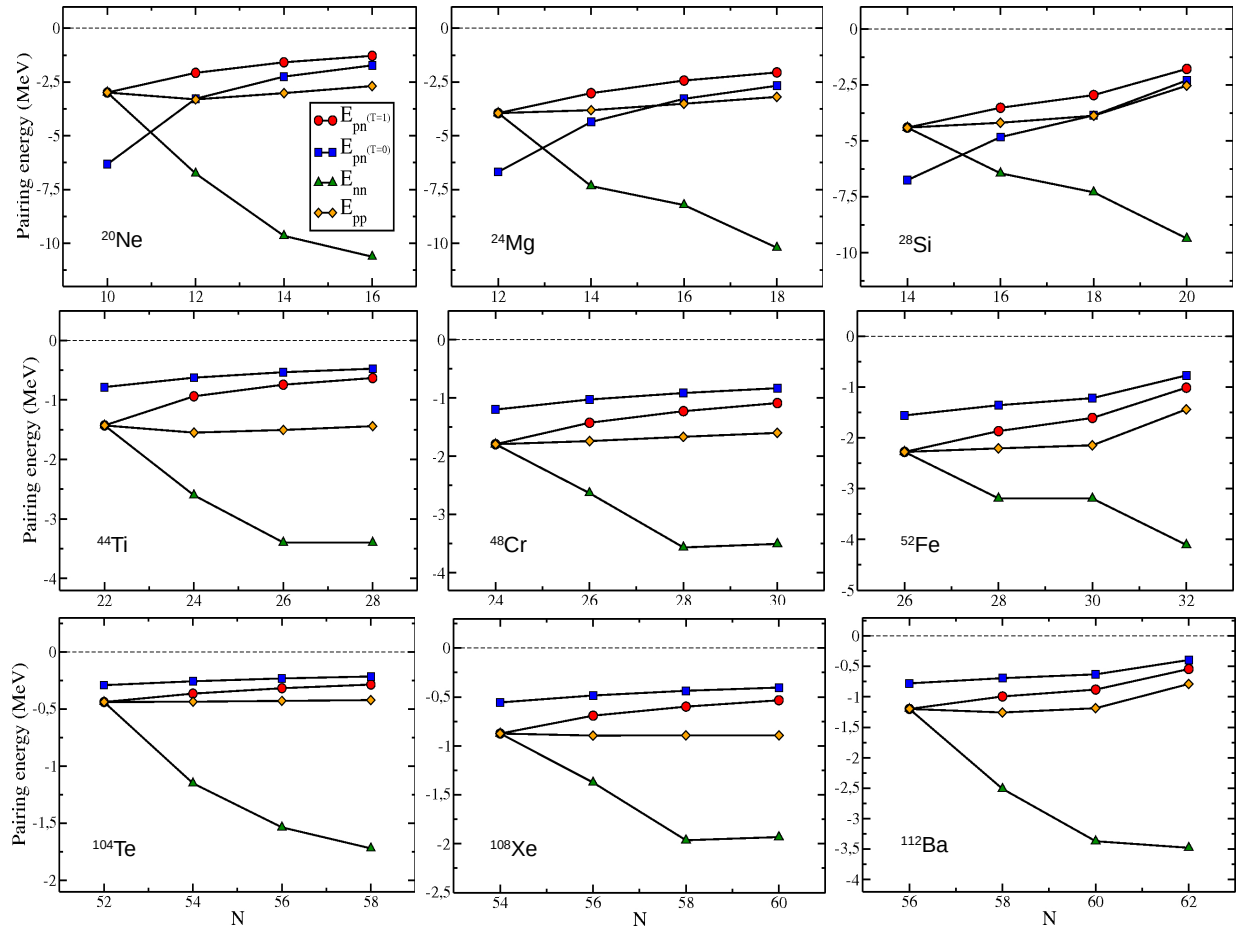}

\end{center}
\caption{Pairing energies, defined by Eqs.(3.11-3.12), as a function of neutron number, for various
nuclei. $E^{T}_{pn}$, $E_{nn}$ and $E_{pp}$ denote, respectively, the proton-neutron, neutron-neutron
and proton-proton pairing energies while $T$ is the isospin.}
\label{fig2}
\end{figure}

To illustrate how the pairing correlations are affected by the extra neutrons, in Fig. 2 are  plotted,
for the state-independent interaction,  the average of the isovector and isoscalar pairing forces.
The latter are defined by
\begin{equation}
E_t^{(T=1)} = g^{(T=1)} \sum_{i,j,t} \langle QCM| P_{i,t}^{\dagger} P_{j,t} |QCM \rangle,
\end{equation}
\begin{equation}
E_{pn}^{(T=0)} = g^{(T=0)} \sum_{i,j} \langle QCM| D_{i,0}^{\dagger} D_{j,0} |QCM \rangle.
\end{equation}
The isovector pairing energies corresponding to isospin projections $t=\{1,-1,0\}$
are denoted by $E_{nn}$, $E_{pp}$ and $E_{pn}^{(T=1)}$. As seen in Fig. 2, with the 
chosen parameters are covered two scenarios concerning the proton-neutron pairing energies, 
i.e.,  nuclei with $E_{pn}^{(T=0)} > E_{pn}^{(T=1)}$ and nuclei with $E_{pn}^{(T=0)} < E_{pn}^{(T=1)}$.
As expected, the proton-neutron pairing energies are decreasing when extra neutrons
are added. For the $sd$-shell nuclei $E_{pn}^{(T=0)}$ is decreasing faster 
than $E_{pn}^{(T=1)}$ while for the heavier nuclei the situation is opposite. 
However, although the proton-neutron energies are decreasing, they remain significantly
large, even for the systems with 6 extra neutrons. Similar features are observed for
the zero-range delta  interaction. Thus, in variance to the predictions of 
many BCS-like studies, these calculations show that the isoscalar and isovector 
proton-neutron pairing correlations: (a) coexist together in both $N=Z$ and $N>Z$ nuclei;
(b) do not vanish quickly by adding few extra neutrons pairs.

\section{Summary}

We have discussed the treatment of isovector and isoscalar pairing Hamiltonians
for the $N>Z$ systems, with the valence nucleons moving in the same single-particle orbits .
The ground state of these pairing Hamiltonians is described by a condensate of
quartets to which is appended a condensate built with
the neutron pairs in excess. The validity of this ansatz for the ground state
was checked for nucleons moving  above the cores $^{16}$O, $^{40}$Ca and
$^{100}$Sn, and for two pairing interactions, one state-independent and
the other a state-dependent zero range force. It is shown that the ansatz 
used for the ground state provides correlation energies which are very close
to the results obtained by diagonalizing exactly the pairing Hamiltonian. 
The calculations done in this framework show that the pairing correlations 
remain significant, in both channels, even in the case when 6 extra neutrons
are added to a N=Z nucleus.
 
\begin{acknowledgments}
This work was supported by the Romanian National Authority for Scientific Research,
CNCS UEFISCDI, Projects No. PN-III-P4-ID-PCE-2016-0481.
\end{acknowledgments}


\begin{thebibliography}{99}
\bibitem{frauendorf} 
S. Frauendorf and A. O. Macchiavelli, Prog. Part. Nucl. Phys. {\bf 78}, 24-90 (2014).
\bibitem{sagawa} H. Sagawa, C. L. Bai, and G. Colò, Phys. Scripta {\bf 91}, 083011 (2016). 
\bibitem{satula}  W. Satula and R. Wyss, Phys. Lett. B {\bf 393}, 1 (1997).
\bibitem{terasaki} J. Terasaki, R. Wyss, and P.-H. Heenen, Phys. Lett. B {\bf 437},
1 (1998).
\bibitem{goodman99} A.L. Goodman, Phys. Rev. C {\bf 60}, 014311 (1999).
\bibitem{goodman2001} A. L. Goodman, Phys. Rev. C {\bf 63}, 044325 (2001).
\bibitem{gezerlis} A. Gerzelis, G.-F. Bertsch, Phys. Rev. Lett. {\bf 106}, 252502 (2011).
\bibitem{goodman_review} A. L. Goodman, Adv. Nucl. Phys. {\bf 11}, 263 (1979).
\bibitem{lane} A. M. Lane, Nuclear Theory (W. A. Benjamin Inc., New York, 1964).
\bibitem{flower} B. H. Flowers and M. Vujicic, Nucl. Phys. {\bf 49}, 586 (1963).
\bibitem{bremond} B. Bremond and J. G. Valatin, Nuclear Physics {\bf 41}, 640 (1963).
\bibitem{zelevinsky} R. A. Senkov and V. Zelevinsky, Phys. At. Nucl. {\bf 74}, 1267
(2011).
\bibitem{yamamura} J. Eichler and M. Yamamura, Nucl. Phys. A {\bf 182}, 33 (1972).
\bibitem{dobes} J. Dobes, and S. Pittel, Phys. Rev. C {\bf 57}, 688 (1998).
\bibitem{qcm_nez}  N. Sandulescu, D. Negrea, J. Dukelsky, C. W. Johnson, Phys. Rev. C {\bf 85}, 061303(R) (2012).
\bibitem{qcm_ngz} N. Sandulescu, D. Negrea, C. W. Johnson, Phys. Rev. C {\bf 86}, 041302(R) (2012).
\bibitem{qcm_t0t1} N. Sandulescu, D. Negrea, D. Gambacurta, Phys. Lett. B {\bf 751} 348 (2015).
\bibitem{qcm_odd} D. Negrea, N. Sandulescu, D. Gambacurta, Prog. Theor. Exp. Phys. 073D05, (2017).
\bibitem{bentley} I. Bentley and S. Frauendorf, Phys. Rev. C {\bf 88}, 014322 (2013).
\bibitem{pacearescu} F. Simkovic, C.C. Moustakidis, L. Pacearescu, A. Faessler, Phys. Rev. C {\bf 68}, 
054319 (2003).
\bibitem{sly4} E. Chabanat, et al., Nucl. Phys. A {\bf 623}, 710 (1997).

\end{thebibliography}
\end{document}